\documentstyle[prl,aps,psfig,floats,amsfonts,amssymb]{revtex}
\begin{document}
 \twocolumn[\hsize\textwidth\columnwidth\hsize\csname  @twocolumnfalse\endcsname
%
\draft
\title{Scaling and Universality in the Anisotropic Kondo Model and the 
Dissipative Two--State System}
\author
{T. A. Costi}
\address
{Universit\"{a}t Karlsruhe, Institut f\"{u}r Theorie der Kondensierten
Materie, 76128 Karlsruhe, Germany}
\maketitle
\begin{abstract}
Scaling and universality in the 
Ohmic two--state system is investigated by exploiting the equivalence 
of this model to the anisotropic Kondo model. For the Ohmic 
two--state system, we find universal scaling functions for the specific heat, 
$C_{\alpha}(T)$, static 
susceptibility, $\chi_{\alpha}(T)$, and spin relaxation function 
$S_{\alpha}(\omega)$ depending on the reduced temperature 
$T/\Delta_{r}$ (frequency $\omega/\Delta_{r}$), with $\Delta_{r}$ 
the renormalized tunneling frequency, and uniquely specified
by the dissipation strength $\alpha$\ ($0<\alpha<1$). 
The scaling functions can be used to extract
$\alpha$ and $\Delta_{r}$ in experimental realizations.
\end{abstract}
\vskip2pc]


The low energy effective model for a large number of physical systems 
corresponds to a two--state system coupled to an environment
\cite{leggett.87,weiss.93}. 
Examples include two--level 
atoms coupled to the electromagnetic field in quantum optics,
electron--transfer reactions in biological systems,
and the tunneling of defects in metallic glasses\cite{examples}.
The simplest model for describing such systems is the 
spin--boson Hamiltonian \cite{leggett.87},
\begin{eqnarray}
H_{SB} & = &-\frac{1}{2}\Delta \hat{\sigma}_{x}
+\frac{1}{2}\epsilon\hat{\sigma}_{z}
        +\sum_{i} \omega_{i}(a_{i}^{\dagger}a_{i}+\frac{1}{2})\nonumber\\
        &+&\frac{1}{2}q_{0}\hat{\sigma}_{z}\sum_{i}
\frac{C_{i}}{\sqrt{2m_{i}\omega_{i}}}(a_{i}+a_{i}^{\dagger})\label{eq:SB}.
\end{eqnarray}
Here, the Pauli matrices $\hat{\sigma}_{i},i=x,y,z$ describe the two--level 
system, $\Delta$ is the bare tunneling matrix element between the states 
$\sigma_{z}=\uparrow$ and $\sigma_{z}=\downarrow$ and $\epsilon$ is a bias.
The environment is represented by an infinite set of 
harmonic oscillators (labeled by the index $i$) 
with masses $m_{i}$ and frequency spectrum $\omega_{i}$
linearly coupled to the coordinate $Q=\frac{1}{2}q_{0}\sigma_{z}$ 
with coupling constants $C_{i}$. In this paper we restrict ourselves 
to Ohmic dissipation, for which the environment spectral
function $J(\omega)=\frac{\pi}{2}
\sum_{i}(\frac{C_{i}^{2}}{m_{i}\omega_{i}})
\delta(\omega-\omega_{i})$ takes the form
$J(\omega)=2\pi\alpha\omega$, for $\omega \ll \omega_{c}$, where $\omega_{c}$
is a high energy cut--off and $\alpha$ is a dimensionless coupling constant 
characterizing the strength of the dissipation.
The two dimensionless couplings of the model are 
$\Delta/\omega_{c}$ and $\alpha$: in this paper we consider 
the region of parameter space, $\Delta/\omega_{c}\ll \alpha<1$, 
which includes the interesting case of a non--adiabatic bath.

The Ohmic spin--boson model has been intensively investigated over the last
10 years \cite{leggett.87,weiss.93}. The main interest has been in 
understanding how the environment influences the dynamics of the two--level 
system and in particular how dissipation destroys quantum coherence
\cite{leggett.87,weiss.93,chakravarty.95,costi.96,lesage.96,strong.97}.
Insight has been gained by exploiting the equivalence of the Ohmic two--state 
system to several other models, including the inverse square Ising 
model\cite{chakravarty.95}, the anisotropic Kondo model (AKM)\cite{costi.96}, 
and the resonant level model\cite{vigmann.78,guinea.85,oliveira.81}.
The qualitative picture that has emerged can be summarized as follows. 
There is a renormalized tunneling frequency $\Delta_{r}$, 
which depends on $\alpha$ and $\Delta$, and which decreases monotonically 
with increasing $\alpha$ for fixed $\Delta$. For $\alpha=0$,\ 
$\Delta_{r}=\Delta$ (decoupled system plus bath) and for $\alpha>0$ the 
renormalization of this scale increases dramatically as the dissipation 
strength is increased to 1: $\Delta_{r}/\omega_{c}\sim 
(\Delta/\omega_{c})^{1/1-\alpha}$. 
Between $\alpha=0$ and $\alpha=1$, there is a range of different 
behaviour from coherent oscillations at zero dissipation to damped
oscillations at intermediate dissipation strengths and eventually to incoherent
relaxation at strong dissipation\cite{leggett.87,weiss.93}. 
At $\alpha > \alpha_{c}(\Delta)\approx 1$ there is a ``localization'' 
transition at $T=0$ corresponding to a vanishing renormalized
tunneling frequency $\Delta_{r}=0$ \cite{bray.82}. 
Here we address one aspect which has not been dealt with in a unified way in the
literature, namely the meaning of universality and scaling in these
models, in particular for physical properties, such as 
thermodynamic and dynamical quantities. We 
apply Wilson's numerical renormalization group (NRG) method to the 
AKM to calculate the specific heat, 
static susceptibility and dynamical susceptibility.
The equivalence of the two models then allows us to discuss 
scaling and universality in the dissipative two--state system. 

The AKM\cite{anderson.69} is given by
\begin{eqnarray}
H &=& \sum_{k,\sigma} \epsilon_{k}c_{k\sigma}^{\dagger}c_{k\sigma} + 
\frac{J_{\perp}}{2}\sum_{kk'}
        (c_{k\uparrow}^{\dagger}c_{k'\downarrow}S^{-} +
         c_{k\downarrow}^{\dagger}c_{k'\uparrow}S^{+})\nonumber\\
  &+& \frac{J_{\parallel}}{2}\sum_{kk'}
         (c_{k\uparrow}^{\dagger}c_{k'\uparrow} -
          c_{k\downarrow}^{\dagger}c_{k'\downarrow})S^{z} 
+ g\mu_{B}hS_{z},\label{eq:AKM}
\end{eqnarray}
where the first term represents non--interacting conduction electrons and the
second and third terms represent an exchange interaction between a localized
spin $1/2$ and the conduction electrons with strength 
$J_{\perp},J_{\parallel}$. 
A local magnetic field, $h$, coupling only 
to the impurity spin in the Kondo model (the last term in Eq.~\ref{eq:AKM}) 
corresponds to a finite bias, $\epsilon$, in 
the spin--boson model. The correspondence 
between $H$ and $H_{SB}$, established via bosonization\cite{guinea.85}, 
implies $\epsilon=g\mu_{B}h$, 
$\frac{\Delta}{\omega_{c}}= \rho_{0} J_{\perp}$ and 
$\alpha=(1+ \frac{2 \delta}{ \pi})^{2}$, where
$\tan{\delta}= -\frac{ \pi \rho_{0} J_{\parallel}}{4}$. $\delta$ is the phase 
shift for scattering of electrons from a potential $J_{\parallel}/4$ and 
$\rho_{0}=1/2D_{0}$ is the conduction electron density of states per
spin at the Fermi level for a flat band of width $2D_{0}$ 
\cite{leggett.87,costi.96,guinea.85}. 
We choose $\omega_{c}=2D_{0}$ so that 
$\Delta = J_{\perp}$ and measure all energies relative 
to $D_{0}=1$. Since we are interested in describing the 
Ohmic two--state system for $\Delta/\omega_{c}\ll \alpha < 1$, this
requires in the AKM that 
$\rho_{0}J_{\perp}\ll 1 - \rho_{0}J_{\parallel}$ for 
$\rho_{0}J_{\parallel}\ll 1$ ($\alpha\rightarrow 1^{-}$ case) 
and $\rho_{0}J_{\perp}\ll
1/(\rho_{0}J_{\parallel})^{2}$ for $\rho_{0}J_{\parallel}\gg 1$ 
($\alpha\rightarrow 0$ case). The AKM for 
$\rho_{0}J_{\perp}>\rho_{0}J_{\parallel}$ will be dealt with elsewhere
so in effect we consider only 
$\rho_{0}J_{\perp} \ll min(\rho_{0}J_{\parallel},\alpha)$.

We solve the AKM using Wilson's NRG method 
\cite{wilson.75+kww.80}. In this procedure, 
a separation of energy scales is made by
introducing a logarithmic mesh of $k$ points $k_{n}=\Lambda^{-n},\; \Lambda>1$,
and transforming the $c_{k\sigma}$ to a basis 
of Wannier states $f_{n\sigma}$\cite{wilson.75+kww.80} at the impurity, with 
$f_{0\sigma}=\sum_{k}c_{k\sigma}$, such that
$H_{c}=\sum_{k\mu}\epsilon_{k\mu}c_{k\mu}^{\dagger}c_{k\mu}$ is
tridiagonal in k--space, i.e. 
$H_{c}\rightarrow \sum_{\mu}\sum_{n=0}^{\infty}\Lambda^{-n/2}
(f_{n+1\mu}^{\dagger}f_{n\mu}+ h.c.)$.
The Hamiltonian (\ref{eq:AKM}) in the new basis is now diagonalized
by defining a sequence of finite size 
Hamiltonians $H_{N}$ containing the first $N$ Wannier states together with
the impurity. One diagonalizes the rescaled Hamiltonians 
$\bar{H}_{N}=\Lambda^{\frac{N-1}{2}}H_{N}$ which satisfy the recursion
relation
$\bar{H}_{N+1} = \Lambda^{1/2}\bar{H}_{N} + 
\sum_{\mu}(f_{N+1\mu}^{\dagger}f_{N\mu}+h.c.)$.
This gives the excitations and eigenstates at a corresponding
set of energy scales $\omega_{N}$ defined by 
$\omega_{N}=\Lambda^{-\frac{N-1}{2}}$ and allows the calculation of
dynamic quantities at frequencies $\omega\sim \omega_{N}$ and 
thermodynamic quantities at temperatures $k_{B}T_{N}\sim \omega_{N}$. 
For example, the Fourier transform of 
$\chi(t,T)=-i\theta(t)\langle[S_{z}(t),S_{z}(0)]\rangle$,
is given by $\chi(\omega,T) = \chi'(\omega,T) + i\chi''(\omega,T) = 
\frac{1}{Z_{N}}\sum_{m,n}|M_{m,n}^{N}|^{2} 
\frac{e^{-\beta\epsilon_{m}}- e^{-\beta\epsilon_{n}}}{
\omega+i0-(\epsilon_{m}-\epsilon_{n})}$,
where $\epsilon_{m},\epsilon_{n}$ are excitations of $H_{N}$, 
$Z_{N}(T)$ the partition function of $H_{N}$, and 
$M_{m,n}^{N}=\langle m|S_{z}|n\rangle_{N}$. 

Specifically, for the AKM we calculate: (a) the $T=0$ relaxation function
$S(\omega)=-\frac{1}{\pi}\frac{{\chi''(\omega+i\delta)}}{\omega}$; (b) the 
impurity specific heat $C(T)=-\partial^{2}F_{imp}/\partial\ T^{2}$, where 
the impurity free energy is given by $F_{imp}(T)=-k_{B}T\ln Z/Z_{0}$ 
and $Z_{0}$ is the conduction electron partition function; (c) the local
static susceptibility $\chi'(\omega=0,T)$ corresponding to setting the 
$g$ factor of the conduction electrons to zero. Under the equivalence, 
the operator $\sigma_{z}/2$ of the spin--boson model translates to 
$S_{z}$ in the AKM, so $S(\omega)$ gives the relaxation function 
for the spin--boson problem. We extract the local static susceptibility, 
$\chi_{sb}(T)=-(1/\beta)(\partial^{2} \ln Z_{sb}/\partial \epsilon^{2})_{\epsilon=0}$, 
of the Ohmic spin--boson model from $\chi'(\omega=0,T)$ at finite $T$.
Finally, the decoupling of spin and charge degrees of freedom in the 
AKM allows identification of 
$F_{imp}$ as the free energy of the spin--boson model (with bath 
contribution subtracted) and of $C(T)$ as the corresponding specific heat.

\begin{figure}[t]
\centerline{\psfig{figure=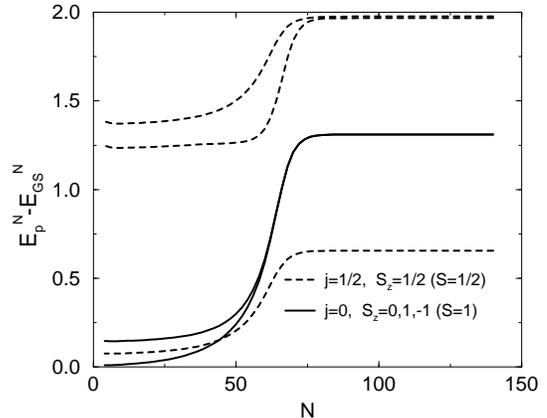,width=8.0cm,angle=0}}
\vspace{0.1cm}
\caption{
The lowest rescaled energy levels 
for even $N$ iterations for $J_{\parallel}=0.443$ 
and $J_{\perp}=0.010$\protect\cite{costi.96} 
corresponding to $\Delta=0.01$ and $\alpha=0.8$ in the 
spin--boson model. The energy levels are labeled by the
conserved quantum numbers, total pseudo--spin $j$ \protect\cite{jones.88}
and total $z$--component of spin $S_z$. There is a crossover 
to the strong coupling fixed point at iteration $N_{c}$ corresponding 
to $\Lambda^{ -(N_{c}-1)/{2} } \approx \Delta_{r}=T_K$.
Spin rotational invariance is restored at low energies 
(e.g., the $j=0$ states with $S_{z}=0$ 
and $S_{z}=\pm 1$ become degenerate), 
so the states at the strong coupling fixed 
point can be labeled by total spin $S$ as indicated.
}
\label{flow-diagram}
\end{figure}
The energy level flow diagram for some low lying rescaled energy states 
is shown in Fig.\ (\ref{flow-diagram}).
We see that spin--rotational invariance, which
is broken for $J_{\perp}\ne J_{\parallel}$ at high energies, 
is restored below the low energy scale of the model, the Kondo temperature 
$T_{K}(J_{\perp},J_{\parallel})$ \cite{tsvelick.83}, leading to the well known
isotropic strong coupling fixed point at low energies (e.g. the lowest single 
particle states in Fig.\ (\ref{flow-diagram}), $\eta_{1}=0.6555, 
\eta_{2}=1.976$ agree with the $\Lambda=2$ results of \cite{wilson.75+kww.80}).
A detailed analysis\cite{costi.97b} gives 
$T_{K}\sim (\Delta/\omega_{c})^{1/(1-\alpha)}$ 
with $\Delta, \alpha$ related
to $J_{\perp},J_{\parallel}$ as above and $J_{\perp}\ll J_{\parallel}$ and a
prefactor also depending on $\alpha$, i.e., 
$T_{K}$ has the same dependence on $\alpha$
as the low energy scale $\Delta_{r}$.
The flow to the isotropic strong coupling fixed point 
holds for any initial anisotropy, corresponding to $0<\alpha<1$, with
the flow being {\em universal} for each $\alpha$\cite{costi.97b}: the
energy levels for fixed $\alpha$ and different $\Delta/\omega_{c}\ll \alpha$ 
may be shifted onto each other by a translation in $N$ 
\cite{special-cases}, except for a 
small ``transient'' region $N\sim 0$--$10$ corresponding to high energies
$\omega \gg \Delta_{r}$ near the cut--off $\omega_{c}$. 
We see that the energy level flow is uniquely specified by two parameters, 
$\Delta_{r}$ (equivalently $T_{K}$) which sets the crossover scale in 
Fig.\ (\ref{flow-diagram}), and the dissipation strength 
$\alpha$ (or equivalently the dimensionless initial coupling constant 
$\rho_{0}J_{\parallel}$). The universal flow of energy levels 
is the origin of the scaling, for fixed $\alpha$ and arbitrary 
$\Delta/\omega_{c} \ll \alpha$, in the thermodynamic quantities we 
discuss below \cite{note-ba}.
\begin{figure}[h]
\centerline{\psfig{figure=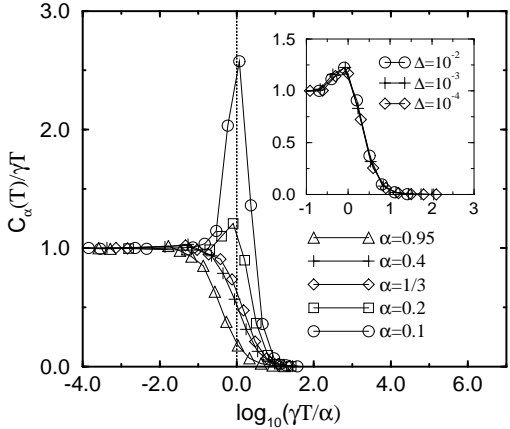,width=8.0cm,angle=0}}
\vspace{0.1cm}
\caption{
Universal specific heat curves, $\frac{C_{\alpha}(T)}{\gamma T}$, 
for the Ohmic two--state system for $0<\alpha<1$. 
$\gamma=\lim_{T\rightarrow 0}C(T)/T\sim \alpha/\Delta_{r}$ 
is extracted from the fixed point analysis\protect\cite{costi.97b}. The
symbols represent the temperatures at which the numerical second derivative
of $F_{imp}$ has been calculated. The inset is for $\alpha=0.2$.
}
\label{specific-heat}
\end{figure}

The inset to Fig.\ (\ref{specific-heat}) shows that the specific heat curves 
for $\alpha=0.2$ and several values of $\Delta$ all scale onto a universal
curve corresponding to $\alpha=0.2$. For different $\alpha$ one obtains distinct
universal curves, $C_{\alpha}(T)$, Fig.\ (\ref{specific-heat}).
Scaling is valid for all temperatures in the range $k_{B}T \ll D_{0}$, 
not only at low temperatures $k_{B}T\ll \Delta_{r}=T_{K}$. 
In Fig.\ (\ref{specific-heat}), and throughout this paper, we
scale the temperature by $\alpha/\gamma  = 3\Delta_{r}/\pi^{2}k_{B}^{2}$. This 
follows from an exact result for the Wilson ratio \cite{vigmann.78,sassetti.90}
discussed below together with our {\em definition} $\chi_{sb}(0)=1/2\Delta_{r}$.
The specific heat is linear in 
temperature for $k_{B}T\ll \Delta_{r}$ and $0<\alpha<1$, 
with a linear coefficient $\gamma$ in good agreement with values extracted 
from an analysis of the strong coupling fixed point
\cite{costi.97b}. This is also expected from the Fermi liquid 
groundstate of the model. We note that the $T^{3}$ coefficient of the 
specific heat changes sign close to $\alpha=1/3$, 
corresponding to the appearance, for weak dissipation, of damped 
oscillations at frequency $\Delta_{r}$ \cite{costi.96,lesage.96}.
The peak in $C(T)/T$ can be taken as a signature of a 
two--level system weakly coupled to bosonic excitations - 
it is absent for strong Ohmic dissipation. The strong dependence of $C(T)/T$
on $\alpha$ for weak dissipation is also seen in other quantities, such
as in the spin response\cite{costi.96}.

\begin{figure}[h]
\centerline{\psfig{figure=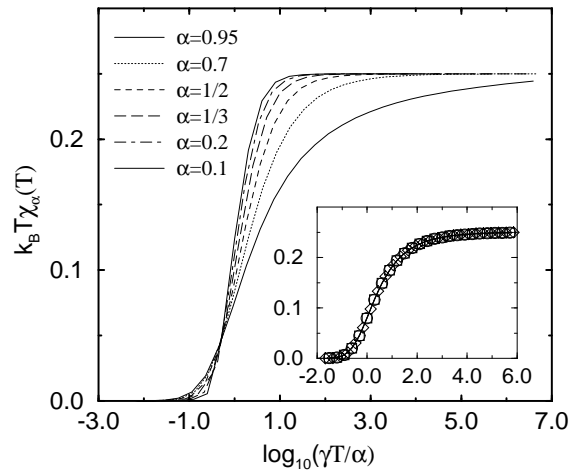,width=8.0cm,angle=0}}
\vspace{0.1cm}
\caption{
Universal curves for the static susceptibility of the Ohmic two--state
system, $k_{B}T\chi_{\alpha}(T)$, for $0<\alpha<1$, with 
$\gamma\sim \alpha/\Delta_{r}$ as in \protect{Fig. \ref{specific-heat}}. The
inset is for $\alpha=0.8$ and $\Delta=0.005$ ($\circ$), $\Delta=0.05$ 
($\square$) and $\Delta=0.01$ ($\diamond$).
}
\label{static-susceptibility}
\end{figure}

The universal curves for the static susceptibility, $\chi_{\alpha}(T)$,
parametrized by the dimensionless dissipation
strength, $\alpha$, are shown in Fig.\ (\ref{static-susceptibility}). 
The inset shows that curves with the same $\alpha=0.8$
and different $\Delta$ scale onto the same curve for $k_{B}T \ll D_{0}$.
The thermodynamic calculation for $\chi_{\alpha}$ becomes
inaccurate at low temperatures, $k_{B}T \ll \Delta_{r}$ 
(discussed in detail in \cite{costi.96,costi.97b}), 
and we have to resort to an analysis about the strong coupling fixed point. 
This yields a finite susceptibility at $T=0$ which is accurate to 
within 1\% for $0<\alpha<1$ (see Table~\ref{table1} below and \cite{costi.97b}). 
At high temperatures, $\Delta_{r}\ll k_{B}T \ll D_{0}$, 
there is a dramatic difference 
in the approach of $k_{B}T\chi_{\alpha}(T)$ to its free spin value of $1/4$ 
between the cases of weak and strong dissipation. In the former, the free spin 
value is reached very rapidly on increasing the temperature 
above $\Delta_{r}$. The logarithmic terms characterizing the slow approach of
$k_{B}T\chi_{\alpha}(T)$ to the free spin value in the Kondo case 
are small for weak dissipation and only set in when
$\alpha\rightarrow 1^{-}$. In this limit and for $\Delta/\omega_{c}\ll 1$, the
scaling functions for the Kondo problem are recovered for the specific heat
and static susceptibility \cite{costi.97b}.

A universal Wilson ratio for the AKM\cite{vigmann.78} has been 
proven also for the Ohmic and non--Ohmic
spin--boson models \cite{sassetti.90}. For the Ohmic case, 
the Wilson ratio for the spin--boson model, 
$R_{sb}=\lim_{T\rightarrow 0}\frac{4}{3}\frac{\pi^{2}k_{B}^{2}}{(g\mu_{B})^{2}}\frac{T\chi_{sb}}{C}
=2/\alpha$ \cite{sassetti.90}. This is related to the Wilson ratio, 
$R_{akm}=\lim_{T\rightarrow 0}\frac{4}{3}\frac{\pi^{2}k_{B}^{2}}{(g\mu_{B})^{2}}\frac{T\chi_{akm}}{C}=2$, for the AKM\cite{vigmann.78} by $R_{sb}=R_{akm}/\alpha$
since $\chi_{sb}=\chi_{akm}/\alpha$\cite{vigmann.78} and 
$\chi_{akm}$ is the susceptibility of the AKM (with a $g$ factor of $2$ for the 
conduction electrons). Table~\ref{table1} 
shows that $R_{akm}=2$ ($R_{sb}=2/\alpha)$ is recovered
for $\Delta/\omega_{c} \ll \alpha$ and $0<\alpha< 1$. 
\narrowtext
\protect\begin{table}
\caption{
$\gamma$, $\chi_{akm}$ and $R_{akm}$, extracted from the fixed point analysis. 
}
\label{table1}
\begin{tabular}{cccccc}
$\alpha$ 
& $\Delta=J_{\perp}$ 
& $\chi_{akm}$
& $\gamma$ 
& $R_{akm}$\\
\tableline
$10^{-3}$ & $10^{-6} $ & $489.2$  &  $3214.6$  & $2.003$ &\\
$10^{-3}$ & $10^{-4} $ & $4.78$  &  $30.76$  & $2.04$ &\\
$1/3$     & $0.01$      & $275.1$  &  $1806.1$  & $2.005$ &\\
$1/3$     & $0.1$     & $8.61$  &  $56.4$  & $2.01$ &\\
$0.7$     & $0.01$     & $1.90\times 10^{6}$ &  $1.25\times 10^{7}$ & $2.002$ &\\
$0.9$     & $0.1$      & $1.81\times 10^{6}$ &  $1.19\times 10^{7}$ & $2.003$ \\
\end{tabular}
\end{table}

\begin{figure}[h]
\centerline{\psfig{figure=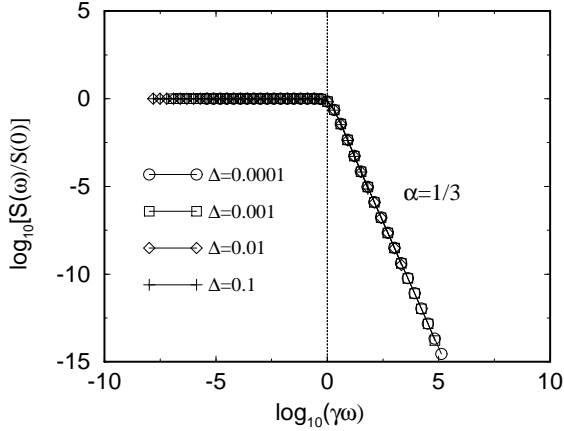,width=8.0cm,angle=0}}
\vspace{0.1cm}
\caption{
Universal curve for the relaxation function $S_{\alpha}(\omega)$.
}
\label{relaxation-function}
\end{figure}

Dynamical quantities also show the universality discussed above for 
thermodynamic quantities. The $T=0$ relaxation functions, $S(\omega)$,
for dissipation strength $0<\alpha<1$ have been given in 
\cite{costi.96}. A detailed analysis shows that these are universal functions
of $\omega/\Delta_{r}$ parametrized by $\alpha$: 
$S=S_{\alpha}(\omega/\Delta_{r})$. The case of
$\alpha=1/3$, corresponding to the crossover between damped oscillations
and incoherent relaxation \cite{costi.96,lesage.96,egger.97}, 
is shown in Fig.\ (\ref{relaxation-function}). 
Scaling in $S_{\alpha}(\omega)$ extends to all 
frequencies $\omega \ll D_{0}$ and is not
restricted to $\omega \lesssim \Delta_{r}$. 
At high frequencies, $\omega \gg \Delta_{r}$, we obtain
$S_{\alpha}(\omega)\sim \omega^{-(4-2\alpha)}$, 
thus $\chi''(\omega)\sim \omega^{-(3-2\alpha)}$ and 
$C_{s}(t)\equiv\langle[\sigma_{z}(t),\sigma_{z}(0)]_{-}\rangle
\sim 1 -ct^{2(1-\alpha)}$ for 
$D_{0}^{-1}\ll t\ll \Delta_{r}^{-1}$ 
with the $\alpha$ dependent exponents being accurate to within
$0.1\%$ for $0<\alpha<1$. These results agree with short--time approximations 
\cite{leggett.87} and perturbative methods\cite{guinea.85} in this limit. 
More importantly, they indicate, that scaling, in the sense
discussed in this paper, can only be expected for fixed $\alpha$.
At low frequencies, $\omega \ll \Delta_{r}$, the Fermi liquid behaviour of 
$\chi''(\omega,T=0)\sim \alpha\omega$,\ gives $C_{s}(t)\sim -\alpha/t^{2}$, 
with an $\alpha$ independent {\em exponent} for $t \gg 
1/\Delta_{r}$\cite{chakravarty.95,costi.96,sassetti.90}.

Deviations from scaling, starting at high temperatures and frequencies, 
set in on increasing $\Delta/\omega_{c}$ for our
finite bandwidth model. 
The scaling discussed here is valid for $0<\alpha <1$ 
($0<J_{\parallel}<\infty$) as long as $\Delta/\omega_{c}$ remains 
the smallest bare energy scale.

The scaling and universality discussed above can be useful in 
interpreting experiments on dissipative two--state systems. 
The dissipation strength $\alpha$ can be determined
by fitting the data for some quantity, such as $S(\omega)$, to the
appropriate universal scaling function $S_{\alpha}(\omega)$. 
The low energy scale, $\Delta_{r}$, can be extracted from the
low energy/temperature behaviour (e.g. from $S_{\alpha}(\omega)$ by 
using the generalized Shiba relation\cite{sassetti.90} 
$S_{\alpha}(0)=2\alpha[\chi_{sb}(T=0)]^{2}$ and $\chi_{sb}(T=0)=1/2\Delta_{r}$). 
In practice, this may be difficult for strong dissipation 
$\alpha\geq 1/3$ since the scaling functions
for different $\alpha$ will differ appreciably only 
for $k_{B}T,\omega > \Delta_{r}$. In this case an alternative is
to extract $\alpha$ and $\Delta_{r}$ from 
$\gamma=\pi^{2}k_{B}^{2}\alpha/3\Delta_{r}$ and 
$\chi_{sb} = 1/2\Delta_{r}$. $\alpha$ and $\Delta_{r}$ can be more easily
deduced for weakly dissipative systems $\alpha < 1/3$, for which the 
scaling functions depend sensitively on $\alpha$, even for $k_{B}T,\omega 
\ll \Delta_{r}$.

To summarize, we have used the equivalence of the Ohmic spin--boson model
to the AKM in order to study universality and scaling in
these models. For anisotropies in the AKM, 
$0<J_{\perp}\ll J_{\parallel}<+\infty$,
corresponding to dissipation strengths $0<\alpha<1$ and bare tunneling
frequencies $\Delta/\omega_{c}\ll \alpha$ 
in the Ohmic spin--boson model, the thermodynamic (dynamic) 
properties of these models are characterized by universal scaling 
functions of $T/\Delta_{r}$ ($\omega/\Delta_{r}$) which 
are {\em distinct} functions for different $\alpha$ and are 
independent of $\Delta$---the latter entering only through $\Delta_{r}$.
As in the Kondo problem, the scaling functions are universal 
with deviations from scaling at high frequencies and temperatures 
arising from finite $\Delta/\omega_{c}$. 
The dissipation strength in the Ohmic two--state system, just as
the anisotropy in the AKM, 
determines the essential physics, in particular the renormalization of the
low energy scale $\Delta_{r}/\omega_{c}\sim (\Delta/\omega_{c})^{1/(1-\alpha)}$
and the form of the scaling functions.
The perspective gained above may also be useful in understanding 
the highly anisotropic multi--channel Kondo models which arise in the
context of single electron devices, and two--level systems in solids interacting
with electrons \cite{cox.97}.  

We acknowledge useful discussions with P. W\"{o}lfle, A. Rosch, C. Roth,
T. Kopp, J. von Delft, J. Kroha and A. E. Ruckenstein. 
This work was supported by the Deutsche Forschungsgemeinschaft and the
German-Israeli Foundation. 


\begin{references}
\bibitem{special-cases} 
This is not true for fixed $\Delta/\omega_{c}\ll \alpha$ and different $\alpha$,
except in the special case of the Kondo limit $\alpha\rightarrow 1^{-}$.
\bibitem{note-ba} This is consistent with the Bethe--Ansatz for the 
AKM \cite{tsvelick.83} where the scaling trajectories are labelled by a
parameter which we can relate to $\alpha$ 
for $0<\alpha<1$ and $\Delta/\omega_{c}\ll \alpha$ \cite{costi.97b}. 
Renormalizations of $\alpha$, entering the scaling functions, are typically
a fraction of a percent (being $O((\Delta/\omega_c)^{2})$) and unimportant
in the regime of interest for the Ohmic two-state system studied in this paper.
\end{references}
\end{document}